\documentclass{PoS}

\usepackage{amsmath}

\title{Current constraints from cosmogenic neutrinos on the fraction of protons in UHECRs}

\ShortTitle{Current constraints from cosmogenic neutrinos on the fraction of protons in UHECRs}

\author{\speaker{Arjen van Vliet}\\
        Deutsches Elektronen-Synchrotron (DESY), Platanenallee 6, 15738 Zeuthen, Germany\\
        E-mail: \email{arjen.van.vliet@desy.de}}

\author{Rafael Alves~Batista\\
        Universidade de S\~ao Paulo, Instituto de Astronomia, Geof\'isica e Ci\^encias Atmosf\'ericas, Rua do Mat\~ao, 1226, 05508-090, S\~ao Paulo-SP, Brazil}

\author{J\"org R. H\"orandel\\
        Radboud University, Department of Astrophysics/IMAPP, P.O. Box 9010, 6500 GL Nijmegen, The Netherlands\\
        NIKHEF, Science Park Amsterdam, 1098 XG Amsterdam, The Netherlands\\
        Vrije Universiteit Brussel, Dept. of Physics and Astronomy, B-1050 Brussels, Belgium}

\abstract{Cosmogenic neutrinos are created when ultra-high-energy cosmic rays (UHECRs) interact with extragalactic photon backgrounds. In general, the expected flux of these cosmogenic neutrinos depends on multiple parameters, describing the sources and propagation of UHECRs. In our recent paper~\cite{vanVliet:2019nse}, we show that a `sweet spot` occurs at a neutrino energy of $E_{\nu} \sim 1$ EeV.  At that energy the flux mainly depends on two parameters, the source evolution and the fraction of protons in UHECRs at Earth for $E_p \gtrsim 30$ EeV. Therefore, with current upper limits on the cosmogenic neutrino flux at $E_{\nu} \sim 1$ EeV and assuming a certain source class, a constraint on the composition of UHECRs can be obtained. This constraint is independent of hadronic interaction models and indicates that the combination of a large proton fraction and a strong source evolution is disfavored. Upcoming neutrino experiments will be able to constrain the fraction of protons in UHECRs even further, and for any realistic model for the evolution of UHECR sources.}

\FullConference{36th International Cosmic Ray Conference -ICRC2019-\\
		July 24th - August 1st, 2019\\
		Madison, WI, U.S.A.}

\begin{document}

\section{Introduction}

With the detection of an astrophysical neutrino flux by IceCube~\cite{Aartsen:2013jdh} a new window to the Universe has opened. 
One possible origin of astrophysical neutrinos detected at Earth is from interactions of cosmic rays with ambient photon fields, so-called cosmogenic neutrinos. 
However, it is not likely that the neutrinos that have been detected by IceCube so far are cosmogenic~\cite{Roulet:2012rv}.
At higher energies (10~PeV~$\lesssim E_{\nu} \lesssim$~100~EeV) both IceCube and the Pierre Auger Collaboration (Auger) provide stringent limits on the expected astrophysical neutrino flux~\cite{Aartsen:2018vtx, Aab:2019auo}. The cosmogenic contribution to the neutrino flux can be constrained by these limits. 

As this cosmogenic neutrino flux is produced by ultra-high-energy cosmic rays (UHECRs) traveling through space, it can be used to indirectly investigate properties of the sources of UHECRs. The expected cosmogenic neutrino flux depends, in general, on a range of different properties of these sources and of the space in between the sources and Earth. The main ingredients needed to compute the cosmogenic neutrino flux are: the maximal energy of cosmic rays at their sources ($E_{\mathrm{max}}$), the spectral index of the energy spectrum for cosmic-ray emission from their sources ($\alpha$)
, the abundance of each nuclear species emitted as UHECRs from their sources, the luminosity, distribution and evolution with redshift ($z$) of the sources of UHECRs and the presence of extragalactic photon backgrounds (cosmic microwave background (CMB) and extragalactic background light (EBL)) and extragalactic magnetic fields in the space in between the sources and our Galaxy.

Another way to constrain the properties of UHECR sources is by looking at UHECR measurements directly. Both the Pierre Auger (Auger) and the Telescope Array (TA) collaborations have provided excellent results on the spectrum~\cite{Fenu:2017hlc, Matthews:2017hlc} and the depth of the shower maximum ($X_\mathrm{max}$) of UHECRs~\cite{Unger:2017fhr, Abbasi:2018nun}. These will improve in the near future with the planned upgrades of both experiments~\cite{Aab:2016vlz, Kido:2017nhz}. The interpretation of the measurements of $X_\mathrm{max}$ in terms of a composition of cosmic rays is, however, not straightforward as it depends on hadronic interaction models which describe the interactions happening in the atmosphere during a cosmic-ray air shower. It would, therefore, be advantageous to get a probe of the composition of UHECRs that does not depend on these hadronic interaction models. In Ref.~\cite{vanVliet:2019nse} we show that the cosmogenic neutrino flux can be used for this purpose.

In, e.g., Refs.~\cite{Aab:2016zth, vanVliet:2017obm, AlvesBatista:2018zui, Heinze:2019jou, Muzio:2019leu} the UHECR spectrum and composition (assuming specific hadronic interaction modes) are fitted and corresponding cosmogenic neutrino fluxes are obtained under the assumptions of a continuous distribution of identical sources and rigidity-dependent maximum energies. In most cases these combined fits find as a best fit a relatively hard spectral index ($\alpha \lesssim 1.3$), a composition dominated by nuclei with a charge $Z \geq 6$ and a relatively low maximum rigidity ($R_\mathrm{max} \equiv E_\mathrm{max} / Z \lesssim 7$ EV). Due to this intermediate to heavy composition and the low $R_\mathrm{max}$ these best-fit scenarios do not have any protons at the highest energies ($E \gtrsim  30$~EeV). Because there are no protons at the highest energies, the expected cosmogenic neutrino flux in these scenarios is so low that they are hardly detectable even with planned neutrino detectors as ARA~\cite{ara2012a}, ARIANNA~\cite{Anker:2019mnx} and GRAND~\cite{Alvarez-Muniz:2018bhp}.

However, that there are no protons at the highest energies in these scenarios is purely a consequence of the assumptions of a continuous distribution of identical sources. This assumption is obviously not what is actually happening in nature. In reality no two sources are identical, let alone all sources in the Universe. Still, this is a reasonable assumption for obtaining indications for the average values of $\alpha$, $R_\mathrm{max}$ and the composition at the sources that dominate the UHECR spectrum and composition at Earth, but it is not enough to obtain a reliable prediction for the expected cosmogenic neutrino flux. If, for example, there are sources which can accelerate cosmic rays to energies much higher than the average $R_\mathrm{max}$, but only give a subdominant contribution to the total UHECR spectrum, we would obtain a certain fraction of protons at Earth at the highest energies and, therefore, also significantly more cosmogenic neutrinos. Having such an additional proton component at the highest energies would only have a minor effect on the expected UHECR spectrum and composition, but would make a big difference in the expected cosmogenic neutrino flux. In fact, Ref.~\cite{Muzio:2019leu} has shown that, for the specific model discussed there, an additional proton component even improves the fit to the UHECR spectrum and composition.

\section{Parameter dependencies of cosmogenic neutrino fluxes}
\label{sec:dep}

In Ref.~\cite{vanVliet:2019nse} we show that the expected cosmogenic neutrino flux at $E_{\nu} \approx 1$~EeV for such an additional proton component is relatively stable. It only depends strongly on $f$, the fraction of protons in UHECRs at Earth at $E_0 = 10^{1.55}$~EeV, and the evolution of the UHECR sources with redshift (parametrized as a function of the source evolution parameter $m$). The contribution to the cosmogenic neutrino flux from heavier nuclei can be safely neglected as long as $f \gtrsim 0.01$ as protons produce many more neutrinos during their propagation (see e.g. Refs.~\cite{Kotera:2010yn, Roulet:2012rv}). Therefore, the current limits on (future measurements of) the neutrino flux at $E_{\nu} \approx 1$~EeV can be used to constrain (determine) which combinations of $f$ and $m$ are viable. 

For this purpose we have parametrized the source evolution (SE), a combination of the evolutions of both the source number density and luminosity, as
\begin{equation}
\mathrm{SE}(z) = 
	\begin{cases}
	  (1 + z)^m & \text{for } m \leq 0 \\
	  (1+z)^m & \text{for } m > 0 \text{ and }  z < 1.5  \\
      2.5^m & \text{for } m > 0 \text{ and } z \geq 1.5
	\end{cases},
  \label{eq:sourceevolution0}
\end{equation}
extending up to a redshift of $z_\mathrm{max} = 4.0$. The contribution to the neutrino flux from sources with $z > z_\mathrm{max}$ is expected to be negligible as there the source evolution with redshift for all typical UHECR source classes is decreasing rapidly.

Besides $f$ and $m$ the expected cosmogenic neutrino flux also depends on $\alpha$, $R_\mathrm{max}$ and the choice of EBL and EGMF models. For the EBL we use the model by Franceschini {\it et al.}~\cite{Franceschini:2008tp}. However, for neutrino energies of $E_{\nu} \gtrsim 0.1$~EeV the CMB is the dominant photon field for neutrino production (see e.g. Ref.~\cite{AlvesBatista:2019rhs}). As we are here mainly interested in the cosmogenic neutrino flux at $E_{\nu} \approx 1$~EeV the choice of EBL model is irrelevant for our purposes. 

Concerning the effects from EGMFs, according to Ref.~\cite{Wittkowski:2018giy} the expected cosmogenic neutrino flux can increase by a factor of a few at $E_{\nu} \approx 1$~EeV, and even by four orders of magnitude at $E_{\nu} \approx 10$~EeV, for the scenarios tested there. However, in all cases shown in Ref.~\cite{Wittkowski:2018giy} the cosmogenic neutrino flux in the range $0.1 \lesssim E_{\nu} \lesssim 1$~EeV is several orders of magnitude larger than the cosmogenic neutrino flux at $E_{\nu} \approx 10$~EeV. Therefore, the relevant energy range for detection of a cosmogenic neutrino flux, for the two cases discussed in that reference, is $0.1 \lesssim E_{\nu} \lesssim 1$~EeV, where the difference between the two curves is much less than at $E_{\nu} \approx 10$~EeV. Additionally, from Ref.~\cite{Wittkowski:2018giy}, it is not clear if the changes to the expected cosmogenic neutrino flux happened due to magnetic-field effects, due to the source distribution or due to the change in fit parameters of composition, spectral index and maximum rigidity. Between the two curves shown in Fig.~1 of Ref.~\cite{Wittkowski:2018giy} not only is the EGMF switched on and off but also all these other parameters changed. These additional changes will also have contributed to the differences seen in Fig.~1 of that paper. Furthermore, the EGMF considered in Ref.~\cite{Wittkowski:2018giy} is stronger than most other recent realistic EGMF models in the literature. The effects of the EGMF on the expected neutrino flux shown in Ref.~\cite{Wittkowski:2018giy} should, therefore, be interpreted as a limiting maximal case. In the scenarios discussed here we neglect any effects of the EGMF. Our predictions of the cosmogenic neutrino flux can, therefore, be considered as lower bounds with the maximum a factor of a few higher in the relevant energy ranges. 

We treat the dependency of the cosmogenic neutrino flux on $\alpha$ and $R_\mathrm{max}$ by computing and showing the results for specific realistic ranges of these two parameters. For $\alpha$ we adopt $1.0 \leq \alpha \leq 3.0$ as full range and for $R_\mathrm{max}$ we choose $1.6 \leq \log(E_\text{max} / \text{EeV}) \leq 5.0$. This range of maximum rigidities is relatively high compared with the low values for $R_\text{max}$ found in recent combined UHECR spectrum and composition fits~\cite{Aab:2016zth, AlvesBatista:2018zui, Heinze:2019jou}. However, the $R_\text{max}$ used here is only for a subdominant additional proton component, not an average for all sources that contribute to the UHECR flux. If $R_\text{max} < 10^{1.6}$~EV for this additional proton component the proton fraction at the highest energies would rapidly go to zero, which would result in something very similar to the best-fit scenarios in Refs.~\cite{Aab:2016zth, AlvesBatista:2018zui, Heinze:2019jou}. It is, however, not unreasonable to expect that at least some of the UHECRs that reach Earth are produced in sources with $R_\text{max} > 10^{1.6}$~EV due to the large variance in intrinsic properties of possible UHECR accelerators across members of the same type of sources.

Besides the full ranges $1.0 \leq \alpha \leq 3.0$ and $1.6 \leq \log(E_\text{max} / \text{EeV}) \leq 5.0$ we also give the results for more restrictive parameter ranges of $1.5 \leq \alpha \leq 3.0$ and $2.0 \leq \log(E_\text{max} / \text{EeV}) \leq 5.0$ and for an even more restrictive parameter range for $\alpha$ of $2.0 \leq \alpha \leq 3.0$. This gives an indication of how the results depend on both the spectral index and the maximum energy, and shows what happens specifically for spectral indices that agree with expectations from Fermi acceleration processes. More detailed investigations of the effects of each of the parameters on the expected spectrum of cosmogenic neutrinos can, for example, be found in Refs.~\cite{Kotera:2010yn,vanVliet:2016dyx,vanVliet:2017obm}.


To compute the expected cosmogenic neutrino fluxes we use the publicly available simulation framework CRPropa 3~\cite{Batista:2016yrx}. We run one-dimensional proton simulations with the CMB and EBL as photon backgrounds and including all relevant interactions; photomeson production, pair production, nuclear decay and adiabatic energy losses due to the expansion of the Universe. At the sources the cosmic rays are assumed to have an injection spectrum of 
\begin{equation}
	\frac{\text{d}N}{\text{d}E} \propto E^{-\alpha} \exp\left(  - \frac{E}{E_\text{max}} \right).
  \label{eq:spec0}
\end{equation}
The sources of the additional proton component are simulated as identical sources distributed according to the evolution with redshift given in Eq.~\ref{eq:sourceevolution0}. The resulting proton spectrum at Earth is then normalized to the Auger spectrum~\cite{Fenu:2017hlc} at $E_0 = 10^{1.55} \; \text{EeV}$ to obtain the $f = 1.0$ case, and the cosmogenic neutrino spectrum is normalized accordingly. To obtain scenarios for $f < 1.0$ both the cosmic-ray and neutrino spectra are straightforwardly shifted by a factor of $f$. The neutrino flux is additionally divided by 3 to get the single-flavor cosmogenic neutrino ($\nu + \bar{\nu}$) spectra assuming a $(\nu_e:\nu_\mu:\nu_\tau) = (1:1:1)$ flavor ratio.

\begin{figure}[tbh]
  \centering
  \includegraphics[width=0.49\columnwidth]{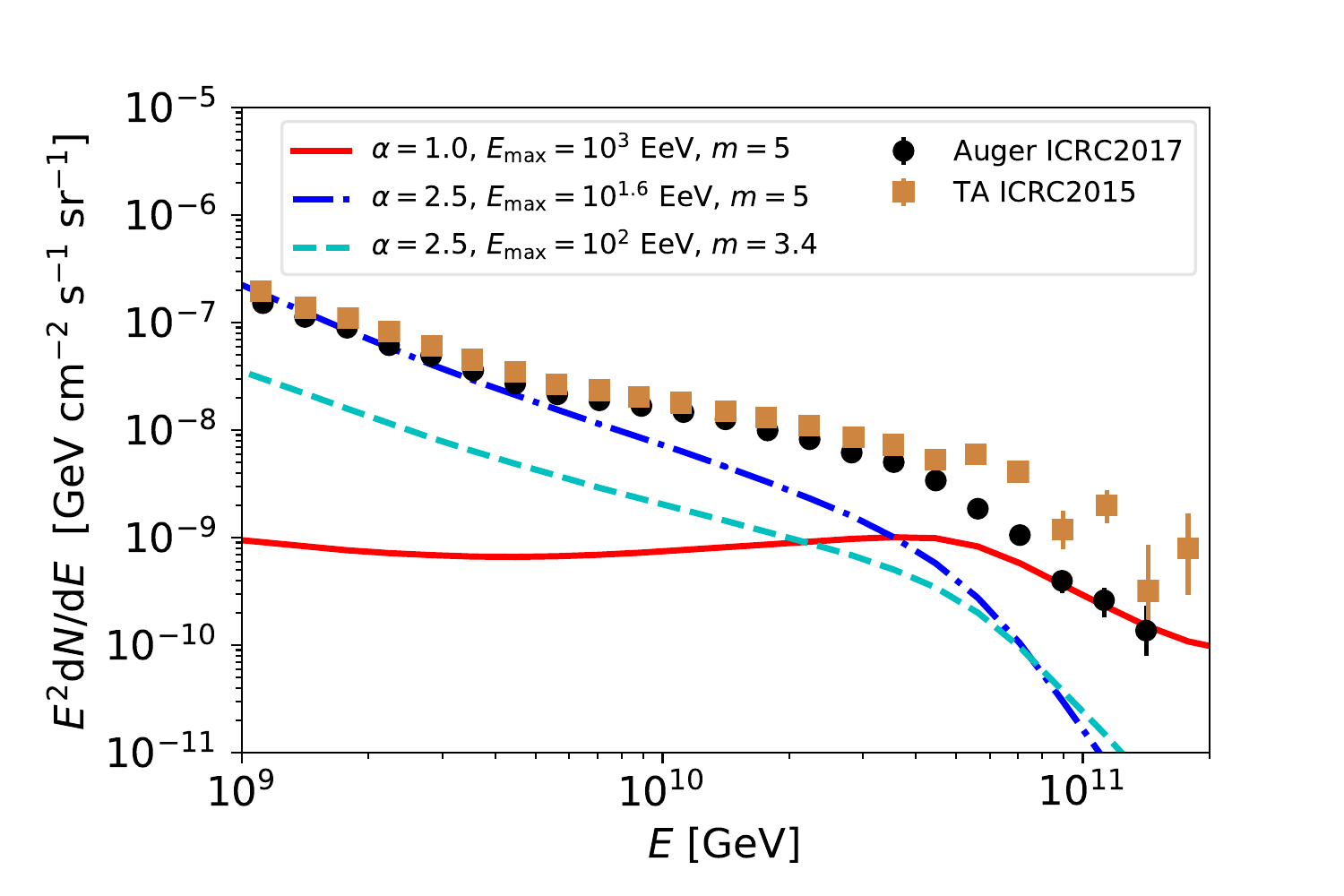}
  \includegraphics[width=0.49\columnwidth]{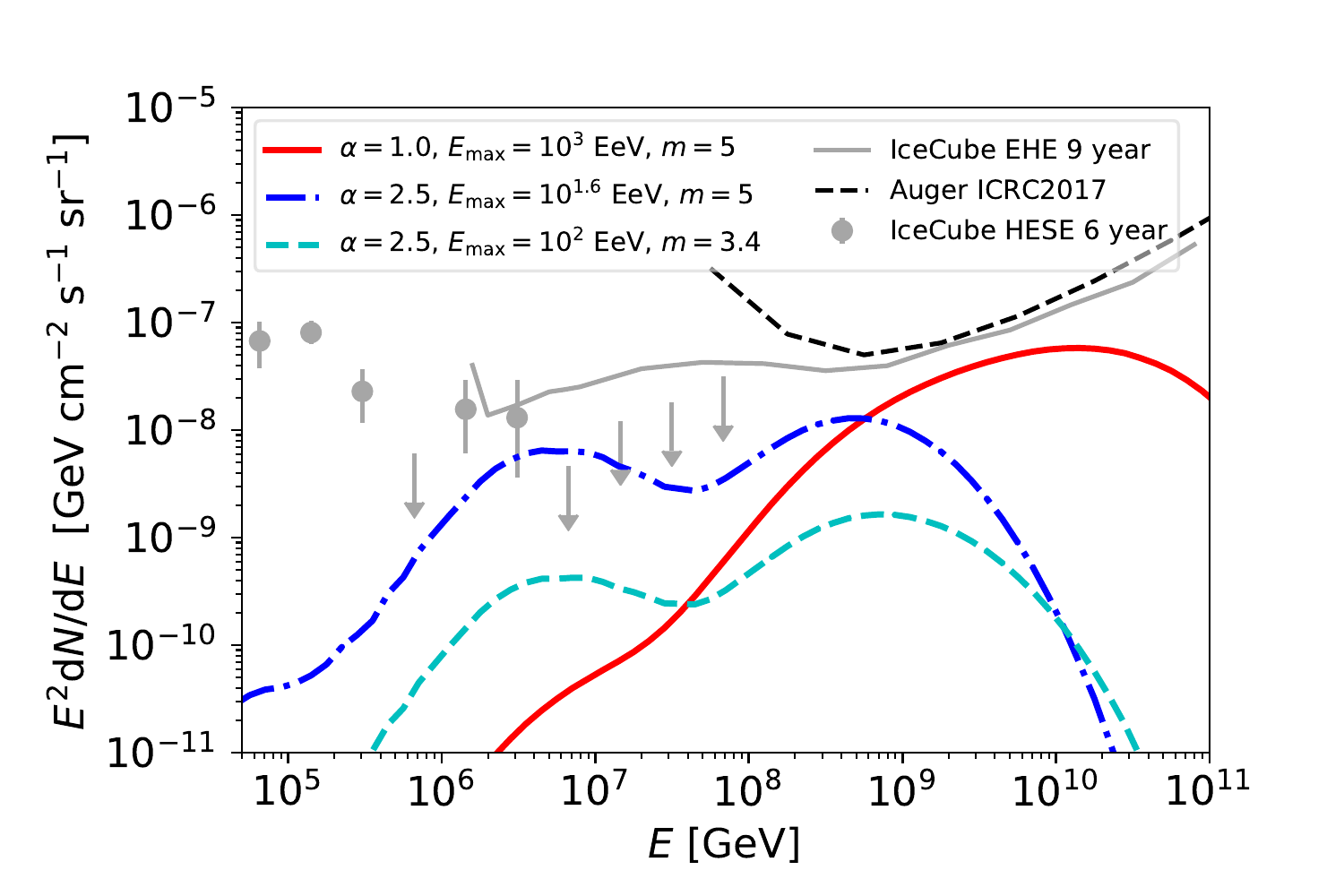}
  \caption{
  {\bf Left} Examples of cosmic-ray spectra for additional proton components that can be added to the full cosmic-ray spectra. The models are normalized to the Auger flux~\cite{Fenu:2017hlc} (black circles) at $E_0 = 10^{1.55} \; \text{EeV}$ and then multiplied by $f$ to get a specific proton fraction at that energy. The red solid line (with $\alpha = 1.0$, $E_\text{max} = 10^3$~EeV and $m = 5.0$) and the dark-blue dashed-dotted line (with $\alpha = 2.5$, $E_\text{max} = 10^{1.6}$~EeV and $m = 5.0$) both have a proton fraction of $f = 0.2$ at $E_0$. The light-blue dashed line (with $\alpha = 2.5$, $E_\text{max} = 10^2$~EeV and $m = 3.4$) has a proton fraction of $f = 0.1$ at $E_0$. For comparison the UHECR spectrum measured by TA~\cite{Matthews:2017hlc} (brown squares) is shown as well. {\bf Right} Expected single-flavor cosmogenic neutrino ($\nu + \bar{\nu}$) fluxes (assuming a $(\nu_e:\nu_\mu:\nu_\tau) = (1:1:1)$ flavor ratio) corresponding to the cosmic-ray spectra in the left figure. To compare the IceCube 6-yr HESE data~\cite{Kopper:2017zzm} and the Auger~\cite{Zas:2017xdj} and IceCube~\cite{Aartsen:2018vtx} differential $90\%$ C.L. upper limits for single-flavor neutrinos and half-energy-decade fluxes are shown as well. 
  }
  \label{fig:maxCRs}
\end{figure}

Examples for proton fluxes that are obtained is this way are given in Fig.~\ref{fig:maxCRs} left, with corresponding cosmogenic neutrino fluxes given in Fig.~\ref{fig:maxCRs} right. Both the red solid line (with $\alpha = 1.0$, $E_\text{max} = 10^3$~EeV and $m = 5.0$) and the dark-blue dashed-dotted line (with $\alpha = 2.5$, $E_\text{max} = 10^{1.6}$~EeV and $m = 5.0$) have a proton fraction of $f = 0.2$ at $E_0$. They are meant as examples of pushed models that do still agree with cosmic-ray measurements and neutrino limits. These two models also have some additional merit. A proton component similar to the dark-blue line could explain the ankle in the UHECR spectrum, while a proton flux similar to the red line might explain the trend towards a lighter composition at the highest energies as observed in Ref.~\cite{Aab:2017cgk}. These two models give the upper range of the shaded area of Fig.~4 left of Ref.~\cite{Ackermann:2019ows} as well as the upper range of the purple shaded area of Fig.~12 left of Ref.~\cite{Anker:2019mnx}.

The light-blue dashed line (with $\alpha = 2.5$, $E_\text{max} = 10^2$~EeV and $m = 3.4$) is a more regular realization of an additional proton component. It has a proton fraction of $f = 0.1$ at $E_0$ and a source evolution close to the star formation rate. This model is the same as the dotted purple line in Fig.~12 left of Ref.~\cite{Anker:2019mnx}. In that figure the potential of ARIANNA is shown for detecting these models.

\section{Results for the proton fraction and source-evolution parameter}

\begin{figure}[tbh]
  \centering
  \includegraphics[width=0.6\columnwidth]{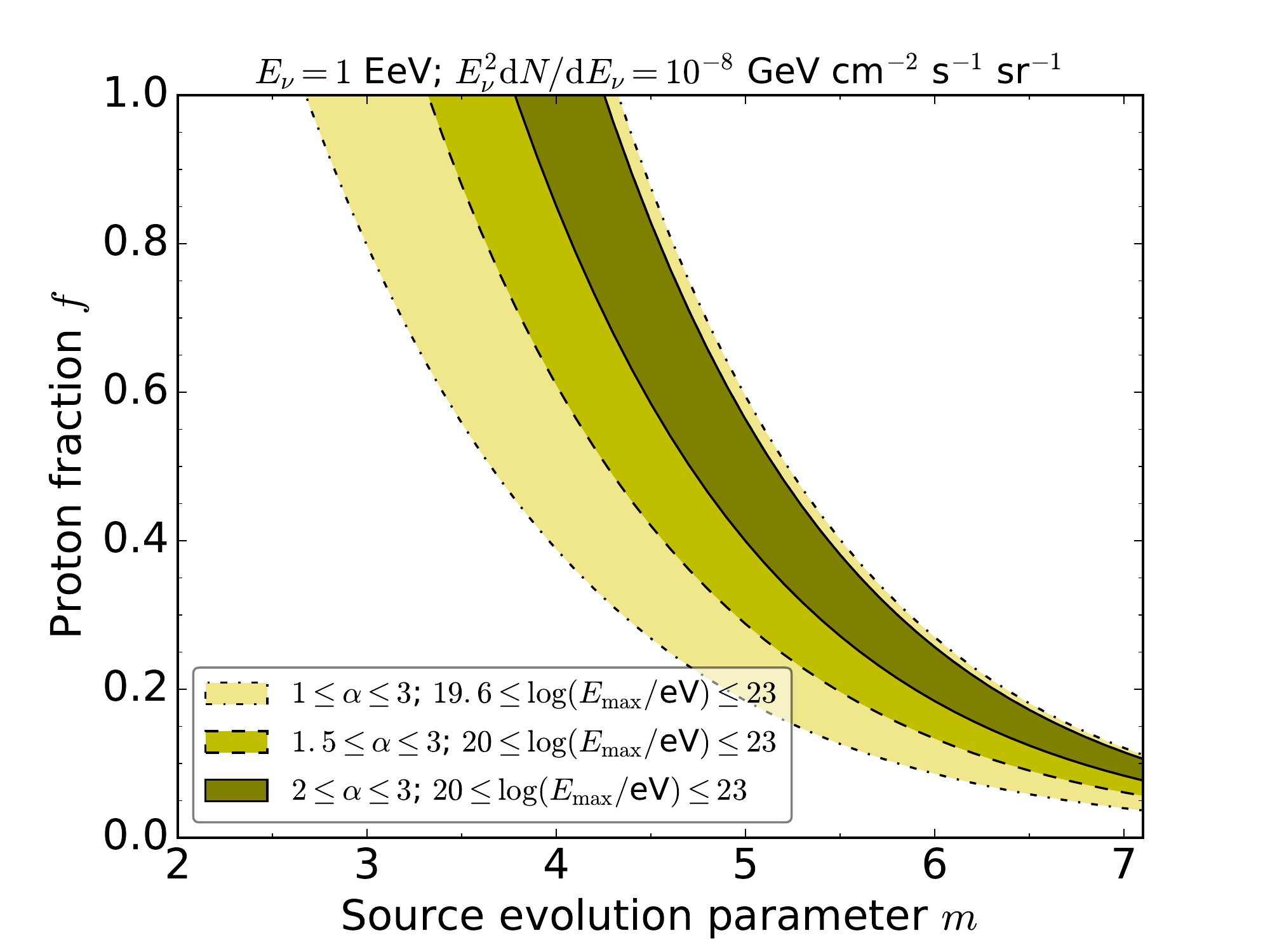}
  \caption{Observable fraction of protons $f$ at ultra-high energies as a function of the source evolution parameter, $m$. A single-flavor flux level of $10^{-8}$~GeV~cm$^{-2}$~s$^{-1}$~sr$^{-1}$ at a neutrino energy of $E_{\nu}=1$~EeV, corresponding roughly to the current sensitivity of IceCube and Auger is assumed. Three shaded areas are shown for different source parameters, from more restrictive (darker) to less restrictive (lighter).
  }
  \label{fig:pFrac}
\end{figure}

As shown in Ref.~\cite{vanVliet:2019nse}, a 'sweet spot' for determining $f$ occurs at a neutrino energy of $E_{\nu} \approx 1$~EeV as there the effect of $\alpha$, $E_\text{max}$ and the EBL on the expected cosmogenic neutrino flux is minimal. So, focusing on this neutrino energy, we can compute all the combinations of $f$ and $m$ that give one specific cosmogenic flux level. This will give an indication for which values of $f$ and $m$ can be expected if a neutrino flux of that level is measured, or which combinations of $f$ and $m$ can be excluded by a neutrino limit at that level. We focus here on a neutrino flux level of $10^{-8}$~GeV~cm$^{-2}$~s$^{-1}$~sr$^{-1}$ at $E_{\nu}=1$~EeV as that corresponds roughly to the current sensitivity of IceCube and Auger. The results for lower flux levels, corresponding to levels that might be reached by ARA, ARIANNA and GRAND in the near future, are given in Ref.~\cite{vanVliet:2019nse}. Fig.~\ref{fig:pFrac} gives the results for a flux level of $10^{-8}$~GeV~cm$^{-2}$~s$^{-1}$~sr$^{-1}$ at $E_{\nu}=1$~EeV for the three different parameter ranges given in Sec.~\ref{sec:dep}. See also Fig.~7 in Ref.~\cite{Aab:2019auo} where constraints on $f$ and $m$ are given using the current Auger limits for $\alpha=2.5$ and $E_\text{max} = 600$~EeV, and Ref.~\cite{Ahlers:2009rf} where older constraints on $f$ are given for two specific source-evolution scenarios.

\section{Conclusions}

Measurements of the cosmogenic neutrino flux at a neutrino energy of $E_{\nu}=1$~EeV can be used to determine the combination of the proton fraction in UHECRs and the evolution of the UHECR sources with redshift. In the same way, limits on the neutrino flux at that energy can be used to constrain the combination of those parameters. In Ref.~\cite{vanVliet:2019nse} we suggest ways to break the degeneracy between these parameters, namely by choosing a specific source type as prior, or by combining the cosmogenic neutrino measurements with measurements of the proton fraction by UHECR experiments. Fig.~\ref{fig:pFrac} shows the results for a flux level of $10^{-8}$~GeV~cm$^{-2}$~s$^{-1}$~sr$^{-1}$. Currently Auger and IceCube put their limits at roughly this flux level. The top right part of Fig.~\ref{fig:pFrac} is, therefore, already constrained by current limits, so the combination of a strong source evolution and a large proton fraction is already ruled out. More specifically, $f \lesssim 0.11$ for $m \gtrsim 7.1$, which corresponds roughly to the source evolution of high-luminosity active galactic nuclei (HL AGNs)~\cite{Hasinger:2005sb}. Therefore, if HL AGNs are the sources of UHECRs, there should be $\lesssim 11$\% protons at a cosmic-ray energy of $E_0 = 10^{1.55} \; \text{EeV}$.

\section*{Acknowledgments}

AvV acknowledges financial support from the NWO Astroparticle Physics grant WARP and the European Research Council (ERC) under the European Union's Horizon 2020 research and innovation programme (Grant No. 646623). RAB is supported by grant \#2017/12828-4, S\~ao Paulo Research Foundation (FAPESP).

\end{document}